\documentclass[12pt]{article}
\usepackage{amsmath}
\usepackage{amsfonts,amsthm}
\usepackage{latexsym}
\usepackage{amscd}
\usepackage[T1]{fontenc}
\usepackage{amssymb}
\usepackage{graphicx}
\usepackage{float}
\usepackage[applemac]{inputenc}
\usepackage{dsfont}
\usepackage{multirow}
\usepackage{dcolumn}
\usepackage{rccol}
\usepackage{enumerate}
\usepackage{scalefnt}
\usepackage{booktabs}
\usepackage{subfigure}

\usepackage{textcomp}
\usepackage{array}
\usepackage{supertabular}
\usepackage{hhline}

\makeatletter
\newcommand\arraybslash{\let\\\@arraycr}
\makeatother
\newcolumntype{+}{>{\global\let\currentrowstyle\relax}}
\newcolumntype{^}{>{\currentrowstyle}}

\newlength{\bracewidth}

\def\fudge{\mathchoice{}{}{\mkern.5mu}{\mkern.8mu}}
\def\bbc#1#2{{\rm \mkern#2mu\vbar\mkern-#2mu#1}}
\def\bbb#1{{\rm I\mkern-3.5mu #1}}
\def\bba#1#2{{\rm #1\mkern-#2mu\fudge #1}}
\def\bb#1{{\count4=`#1 \advance\count4by-64 \ifcase\count4\or\bba A{11.5}\or
   \bbb B\or\bbc C{5}\or\bbb D\or\bbb E\or\bbb F \or\bbc G{5}\or\bbb H\or
   \bbb I\or\bbc J{3}\or\bbb K\or\bbb L \or\bbb M\or\bbb N\or\bbc O{5} \or
   \bbb P\or\bbc Q{5}\or\bbb R\or\bbc S{4.2}\or\bba T{10.5}\or\bbc U{5}\or
   \bba V{12}\or\bba W{16.5}\or\bba X{11}\or\bba Y{11.7}\or\bba Z{7.5}\fi}}

\restylefloat{float}

\title{Role of short-term dispersal on the dynamics of Zika virus}
\author{Victor Moreno$^1$, Baltazar Espinoza$^1$,\\
Derdei Bichara$^1$, Susan A. Holechek$^{1,2}$ and Carlos Castillo-Chavez$^1$\\
$^1$ Simon. A. Levin Mathematical, Computational and Modeling Sciences Center,\\
Arizona State University, P.O. Box 873901, Tempe, AZ 85287-3901; \\
 $^2$ Center for Infectious Diseases and Vaccinology, The Biodesign Institute, \\
Arizona State University, P.O. Box 875401, Tempe, AZ 85287-5401\\
}

\date{}

\begin{document}
\maketitle

\begin{abstract}
In November 2015, El Salvador reported their first case of Zika virus (Zv) leading to an explosive outbreak that in just two months had over 6000 suspected cases. Many communities along with national agencies initiated the process to implement control measures that ranged from vector control and the use of repellents to the suggestion of avoiding pregnancies for two years, the latter one, in response to the growing number of microcephaly cases in Brazil. In our study, we explore the impact of short term mobility between two idealized interconnected communities where disparities and violence contribute to the Zv epidemic. Using a Lagrangian modeling approach in a two-patch setting, it is shown via simulations that short term mobility may be beneficial in the control of a Zv outbreak when risk is relative low and patch disparities are not too extreme. However, when the reproductive number is too high, there seems to be no benefits. \\
This paper is dedicated to the inauguration of the Centro de Modelamiento Matem\'{a}tico Carlos Castillo-Ch\'{a}vez at Universidad Francisco Gavidia in San Salvador, El Salvador.\end{abstract}



\section{Introduction}
Zika virus (Zv), an emerging mosquito-borne flavivirus related to yellow fever, dengue, West Nile and Japanese encephalitis \cite{hayes2009zika}, has taken the Americas by a storm. Zv is transmitted primarily by \textit{Aedes aegypti} mosquitoes, which also transmits dengue and chikungunya, are  responsible for huge outbreaks in some regions in Brazil, Colombia, and El Salvador. Zv  cases as of February 9, 2016, according to the CDC \cite{CDC2016e}, have been reported throughout the Caribbean, Mexico as well as in most  South American nations except for Chile, Uruguay, Argentina, Paraguay and Peru. Several states within the United States have reported Zv cases\cite{petersen2016interim} and although it is expected that Zv will be managed within the USA, the possibility of localized Zv outbreaks is not out of the question.\\

Phylogenetic analyses have revealed the existence of two main virus lineages (African and Asian) \cite{faye2014molecular, haddow2012genetic}. Up to date, there are no concise clinical differences between infections with either one of these two lineages. This is in part due to the few human isolates belonging to the African lineage which were mainly obtained from sentinel rhesus in 1947 in Uganda, where it was first discovered during primate and mosquito surveillance for Yellow Fever \cite{dick1952zika}. Its geographic habitat at that time was tied in to narrow equatorial belt running across Africa and into Asia. The African lineage circulated primarily in wild primates and arboreal mosquitoes such as \textit{Aedes africanus};  spillover infections in humans rarely occurred even in areas were it was found to be highly enzootic \cite{musso2015zika,fauci2016zika}. The more geographically expanded Asian lineage seems to have originated from the adaptation of the virus to invade a different vector, \textit{Aedes aegypti}, which unfortunately seems to be perfectly adapted to infect humans \cite{haddow2012genetic, fauci2016zika}. The first human infection was reported in Nigeria in 1954 \cite{macnamara1954zika}. In 2007, Zika moved out of Africa and Asia causing an outbreak in Yap Island in the Federated States of Micronesia \cite{duffy2009zika}, this was followed by a large outbreak in French Polynesia in 2013-2014 \cite{cao2014emerging} and then spreading to New Caledonia, the Cook Islands and Eastern Islands \cite{musso2014rapid}. Some evidence exists regarding Zika spread following chikungunya epizootics and epidemics as African researchers observed this decades ago. This pattern was also present in 2013 when chikungunya pandemic spread from west to east followed by Zika \cite{fauci2016zika}. In early 2015, Zv was detected in Brazil. Phylogenetic analyses of the virus isolated from patients placed the Brazilian strains in the Asian lineage \cite{zanluca2015first}, which has been previously detected during the French Polynesian outbreak \cite{is2014zika}. Since the first detection of Zv in Brazil, the following countries have reported ongoing substantial transmission of Zv in South America: Bolivia, Brazil, Colombia, Ecuador, French Guyana, Guyana, Paraguay, Suriname and Venezuela \cite{CDC2016c}. Several Central America countries are also affected including Costa Rica, El Salvador, Guatemala, Honduras, Nicaragua and Panama \cite{CDC2016d}. The rapid expansion of Zv has led the World Health Organization (WHO) to declare it a public health emergency of international concern \cite{WHO2016}. \\

It has been estimated that about 80\% of persons infected with Zv are asymptomatic \cite{duffy2009zika,CDC2016} and it is known that those with clinical manifestations present dengue-like symptoms that include arthralgia, particularly swelling, mild fever, lymphadenopathy, skin rash, headaches, retro orbital pain and conjunctivitis which normally last for 2-7 days \cite{zanluca2015first,WHO2016, CDC2016}. Due to the similarity in clinical characteristics between dengue, chikungunya and Zv,  the lack of widely distributed Zv-specific tests, the high proportion of asymptomatic individuals,  it may turn out that the number of patients infected with Zv may actually be a lot higher than what it is being reported \cite{fauci2016zika, salvador2015entry}. Moreover, co-infection with dengue and Zv is not uncommon. In fact, it has been previously reported making Zv diagnosis even more difficult\cite{dupont2015co}. \\

The challenges linked to the control of Zv must include the fact that there is no vaccine available, a troublesome situation given the fact that Zv has  been linked recently to potential neurological (microcephaly) and auto-immune (Guillain-Barr\'{e} syndrome) complications. Further the evidence so far points to the  likelihood that Zv can also be sexually transmitted \cite{CDC2016b,WHO2016}.
Education about Zv modes of transmission and ways of preventing transmission are essential in order to halt mosquito growth and thus Zv spread at the community, population, regional, national and global levels. Control measures available are limited and include the use of insect repellents to protect us against mosquito bites  and  sex abstinence or protection while engage in sexual activity  \cite{CDC2016b}. Some countries face immense challenges that if not addressed would make current efforts by  officials to educate the public highly ineffective. And additional challenges that may limit if not stop the use of whatever control or education measures that a city, or nation or region may be able to put in place, are tied in to the effectiveness of public safety, violence and organized crime activity (including gangs). The Latin America and the Caribbean population comprise only  9\% of the global population and yet, it accounts for  33\% of the world's homicides \cite{BID2015}. In this study, we analyze the impact of that restrictions to public safety may have in the  control of Zv. The motivation comes from our interest in addressing the role of violence and insecurity within the context of the Caribbean, particularly El Salvador.
\section{Derivation of the model}
\label{sec:modelderiv}
As specified in the introduction, the objective of this manuscript is to look at the impact of mobility and security on the transmission dynamics of the Zika virus (Zv), in regions where insecurity, violence and resource limitations make it difficult to implement effective intervention efforts. As a first step, we proceed to build a two-patch model, both patches defined by their relationship to security and associated per capita resources. And so, the first patch is defined by low level security, which  makes it difficult to have access and carry out systematic vector control efforts due, for example, to gang activity. The second patch a safe territory, that is, a place where security is high, access to health services is expected, and relative high levels of education are the norm.\\

Building detailed parametrized models that account for all the above factors would require  tremendous amount of data and information and so,  the model would require a large number of parameters, some that have never been measured. Its use as a policy simulation tool would also require impressive amounts of information on, for example, individuals'  daily scheduled activities.  Here, it is assumed, that each patch is made up of individuals all experiencing the same degree of risk to infection. It is also assumed that all individuals are typical representatives of either the high risk (Patch1) or low risk (Patch 2) communities. The level of risk (violence and infection) is incorporated within a single parameter $\hat{\beta}_i, i=1,2$ and so, by definition, we have in general that  $\hat{\beta}_1 \gg  \hat{\beta}_2$. This assumption captures in a rather simplistic way the essence of what we want to address, namely, the fact that we are looking at the dynamics of Zv in a highly heterogeneous world, modeled, for the purpose of exploring its role, in as simple scenario as possible. Here we look at an extreme case where the situation, idealistically defined via two patches; a high and a low risk patch. The dynamics of individuals in both patches, short time scales are incorporated. The analysis is over the duration of a single outbreak.\\

In the rest of this section, we introduce the prototypic model that will be used to model Zv dynamics within a patch. And so, we let  $N_h$ denote the host patch population size interacting with a vector population of size $N_v$.  The transmission process is model through the interactions of the following epidemiological state variables. And so, we let  $S_h$, $E_h$, $I_{h,a}$, $I_{h,s}$ and $R_h$ denote the susceptible, latent, infectious asymptomatic, infectious symptomatic and recovered sub-populations, respectively while  $S_v$, $E_v$ and $I_v$  are used to denote the susceptible, latent and infectious mosquito sub-populations. 
Since the focus is on a single outbreak, we neglect the hosts' demography while assuming that the vector's demography does not change, this is done, by simply assuming that the  birth  and death per capita mosquito rates are the same. New reports point \cite{CDC2016,duffy2009zika} to the identification of an increasing number of asymptomatic  infectious individuals from  ongoing Zv outbreaks. And so, we  consider two classes of infectious $I_{h,a}$ and $I_{h,s}$, asymptomatic and symptomatic individuals. Furthermore, since not much is known about the dynamics of Zv transmission, we assume that $I_{h,a}$ and $I_{h,s}$ individuals are equally infectious and that their periods of infectiousness are roughly the same. This is not a terrible assumption given our current knowledge of Zv epidemiology and the fact that in general Zv infections are not severe. Furthermore, since the infectious process of Zv is similar to dengue we proceed to use parameter estimates for dengue transmission in El Salvador \cite{fauci2016zika, salvador2015entry} as well as our current estimates of the basic reproduction number for Zv obtained from the data on Barranquilla's Colombia current Zv outbreak \cite{towers2016barranquilla}. The selection of model parameters ranges benefited from those estimated from the 2013-2014 French Polynesia outbreak \cite{kucharski2016transmission}. The dynamics of the prototypic single patch system, single epidemic outbreak,  is  modeled via the following nonlinear system of differential equations:
\begin{equation} \label{1Patch}
\left\{\begin{array}{llll}
\dot S_{h}=-b \beta_{vh}S_{h}\frac{I_{v}}{N_{h}}\\
\dot E_{h}=b\beta_{vh}S_h\frac{I_{v}}{N_{h}}-\nu_h E_{h}\\
\dot I_{h,s}=(1-q)\nu_hE_{h}-\gamma_{h} I_{h,s}\\
\dot I_{h,a}=q\nu_hE_{h}-\gamma_{h} I_{h,a}\\
\dot R_{h}=\gamma_{h}(I_{h,s}+I_{h,a})\\
\dot S_{v}=\mu_vN_v-b\beta_{hv}S_{v}\frac{ I_{h,s}+I_{h,a}}{ N_{h}}-\mu_vS_{v}\\
\dot{E}_{v}=b\beta_{hv}S_{v}\frac{ I_{h,s}+I_{h,a}}{ N_{h}}-(\mu_v+\nu_v)E_{v}\\
\dot{I}_{v}=\nu_v E_v-\mu_v I_{v}
\end{array}\right.
\end{equation}
%
%
%
\begin{table}[h!]
  \begin{center}
    \caption{Description of the parameters used in System (\ref{1Patch}).}
    \label{tab:Param}
    \begin{tabular}{cc}
      \cline{1-2}      
      Parameters & Description \\
      \cline{1-2}
$\beta_{vh}$ & Infectiousness of human to mosquitoes\\
$\beta_{hv}$ & Infectiousness of mosquitoes to humans\\
$b_{i}$ &  Biting rate in Patch $i$ \\
$\nu_h$ & Humans' incubation rate\\
$q$ & Fraction of latent that become asymptomatic and infectious\\
$\gamma_i$  & Recovery rate in Patch $i$\\
$p_{ij}$ &  Proportion of time residents of Patch $i$ spend in Patch $j$\\
$\mu_v$  & Vectors' natural mortality rate\\
$\nu_v$ & Vectors' incubation rate\\
      \cline{1-2}
    \end{tabular}
  \end{center}
\end{table}

\begin{figure}[H]
\begin{center}
\includegraphics[scale=0.75]{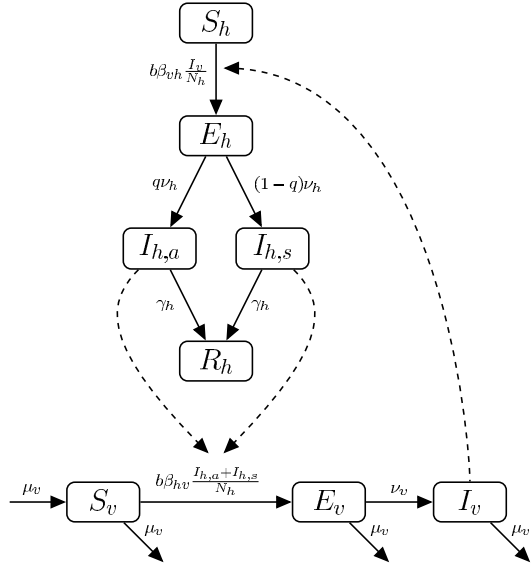}
\caption{Flow diagram of the model}
\label{fig:Flow}
\end{center}
\end{figure}
%
%


The parameters of Model \ref{1Patch} are collected and described in Table \ref{tab:Param} while the flow diagram of the model is provided in Fig \ref{fig:Flow}.  We now proceed to compute the formula for the {\it basic reproduction number} for this prototypic model, that is, the average number of secondary infections generated by a typical infectious individual in a population where nobody has experienced a Zv-infection. That is, we take $S(0) = N_h$, that is, we focus on perturbations of the disease free equilibrium of Model (\ref{1Patch}), that is, on $E_0=(N_h,0,0,0,0)$. And so, using standard approaches, we find that the basic reproduction number  is  given by  
\begin{eqnarray*}\mathcal R_0^2&=&\frac{b^2N_v\beta_{vh}\beta_{hv}\nu_v[(1-q)\gamma_{h,a}+q\gamma_{h,s}]}{N_h\gamma_{h,a}\gamma_{h,s}\mu_v(\mu_v+\nu_v)}\\
& :=& \mathcal R_{0,a}^2+\mathcal R_{0,s}^2,
\end{eqnarray*}
where $\displaystyle\mathcal R_{0,a}^2 =\frac{b^2N_v\beta_{vh}\beta_{hv}\nu_v(1-q)\gamma_{h,a}}{N_h\gamma_{h,a}\gamma_{h,s}\mu_v(\mu_v+\nu_v)}$ the average number of secondary cases produced by infectious asymptomatic during their infectious period whereas $\displaystyle\mathcal R_{0,s}^2=\frac{b^2N_v\beta_{vh}\beta_{hv}\nu_vq\gamma_{h,s}}{N_h\gamma_{h,a}\gamma_{h,s}\mu_v(\mu_v+\nu_v)}$ is the average number of secondary cases produced by a symtptomatic infectious during their infectious period.  We also define the level of risk $\hat{\beta}$ as the product of the biting rate, vector-host ratio and the infectiousness of humans to mosquitoes, $\hat{\beta}=b\beta_{vh} \frac{N_{v}}{N_{h}}$. 
%
%
The dynamics of the single patch model are well known. In short, if $\mathcal R_0^2  \leq 1$, there is no epidemic out break, that is, the proportion of introduced infected individuals, decrease while if   $\mathcal R_0^2  > 1$ we have that the host population experiences an outbreak, that is, the number of cases exceeds the initial size of the introduced infected population at time $t=0$. When $\mathcal{R}_0^2>1$, the population of infected individuals eventually decreases and the disease dies out (single outbreak model). \\

In the next section, a two patch model using a Lagrangian approach found in \cite{bichara2015vector,bichara2015sis} is introduced. And so, individual in Patch $i, i = 1,2$ never loose their residency status. The mobility of patch, residents visiting other patches is modeled by the use of a {\it residence times matrix}. Each entry models the proportion of time that each resident spends in his own patch or as a visitor, every day. We make use of the  $2\times2$ matrix $\mathbb{P}$, with entries $p_{ij}, i, j = 1,2$; $p_{i1} + p_{i,2} = 1, i= 1,2$. Parameters specific to El Salvador are used and the most recent estimates of $\mathcal R_0$ for Zv we explore, via simulations, the consequences of mobility, modeled by the matrix $\mathbb{P}$, the impact that the differences in risk, captured on the assumption $\hat{\beta_1} \gg \hat{\beta_2}$, have on the transmission of Zv by asymptomatic and symptomatic infectious individuals. 

\section{Heterogeneity through residence times}
The role of mobility between two communities, within the same city, living under dramatically distinct health, economic, social and security settings is now explored using as simple model as possible. We need two patches, the inclusion of within- and across-patches vector-host transmission, and the ability of each individual to `move' without loosing, within the model, its place of residence. We consider two highly distinct patches. Patch 1 with access to health facilities, regular security, resources and effective public health policies, while Patch 2 lacks nearly everything.  Within this highly simplified setting, the differences in risk, which depend on host vector ratios, biting rates, use of repellents, access to vector control crews and more, are captured by assuming dramatic differences in transmission; all captured by the single parameter $\hat{\beta}$. It is therefore assumed that $\hat{\beta_1} \gg  \hat{\beta_2}$, where $\hat{\beta_i}$ defines the risk in patch $i, i=1,2$ [high risk Patch 1 and  low risk Patch 2].\\

The two-patch Lagrangian model makes use of a host population stratified by epidemiological classes indexed by residence patch, that is, we  let $S_h$, $E_h$, $I_{h,a}$, $I_{h,s}$ and $R_h$ denote the susceptible, latent, infectious asymptomatic, infectious symptomatic and recovered sub-populations, respectively while  $S_v$, $E_v$ and $I_v$  denote the susceptible, latent and infectious mosquito sub-populations.  Again, $N_h$ denotes the host patch population size and $N_v$  the total vector population in each patch. It is further assumed, a reasonable assumption in the case of {\it Aedes aegypti}, that the the vector does not travel between patches. The single-patch model-parameters are collected and described in Table \ref{tab:Param} while the flow diagram of the single-patch dynamics model, when residents and visitors do not move; that is, when the  $2\times2$ residence times matrix $\mathbb{P}$ is such that  $p_{11} =p_{22} =1$, are captured in Fig \ref{fig:Flow}\\

The   $2\times2$ residence matrix $\mathbb{P}$ with entries $p_{ij}, i, j = 1,2$,  $p_{i1} + p_{i,2} = 1, i= 1,2$, where $p_{ij}$ denotes the proportion of the day that a resident of Patch $i$ spends in Patch $j, i = 1,2$, defines host-mobility, Lagrangian  approach used here. The use of this approach, under the assumption that vectors do not move \cite{bichara2015vector}, has consequences. For example, it leads to the  conclusion that at time $t$, the population on each patch does not necessarily have to be equal to the number of residents in that patch, in other words, the {\it effective population patch size} must account for  residents and visitors at {\it each patch at time $t$}. The specifics of the model are now provided following the work of Bichara et al. \cite{bichara2015vector,bichara2016dynamics}.

\subsection{Two patch model}
Following our recently developed version of residence patch models \cite{bichara2015vector,bichara2016dynamics,bichara2015sis}, leads to the following two patch model where $\mathbb{P} = (p_{ij})$ and $p_{i,j}$ represents the residence time that an individual from Patch $i$ spends in Patch $j; i= 1,2$.

\begin{equation} \label{2Patch}
\left\{\begin{array}{llll}
\dot S_{h,i}=-\beta_{vh}S_{h,i}\sum_{j=1}^{2} b_jp_{ij}\frac{I_{v,j}}{p_{1j}N_{h,1}+p_{2j}N_{h,2}}\\
\dot E_{h,i}=\beta_{vh}S_{h,i}\sum_{j=1}^{2} b_jp_{ij}\frac{I_{v,j}}{p_{1j}N_{h,1}+p_{2j}N_{h,2}}-\nu_{h,i} E_{h,i}\\
\dot I_{h,s,i}=(1-q)\nu_{h,i}E_{h,i}-\gamma_{h,s} I_{h,s,i}\\
\dot I_{h,a,i}=q\nu_{h,i}E_{h,i}-\gamma_{h,a} I_{h,a,i}\\
\dot R_{h,i}=\gamma_{h,s}I_{h,s,i}+\gamma_{h,a}I_{h,a,i}\\
\dot S_{v,i}=\mu_vN_{v,i}-b_i\beta_{hv}S_{v,i}\frac{ \sum_{j=1}^{2}p_{ji}(I_{h,s,j}+I_{h,a,j})}{ \sum_{k=1}^{2}p_{ki}N_{h,k}}-\mu_vS_{v,i}\\
\dot{E}_{v,i}=b_i\beta_{hv}S_{v,i}\frac{ \sum_{j=1}^{2}p_{ji}(I_{h,s,j}+I_{h,a,j})}{ \sum_{k=1}^{2}p_{ki}N_{h,k}}-(\mu_v+\nu_v)E_{v,i}\\
\dot{I}_{v,i}=\nu_v E_{v,i}-\mu_v I_{v,i}
\end{array}\right.
\end{equation}

The basic reproduction of this model is the largest eigenvalue of the following matrix \cite{bichara2015vector},

$$M_1=\left(\begin{array}{cc}
m_{11} & m_{12}\\
m_{21} & m_{22}
\end{array}\right)
$$
where 
\begin{eqnarray*}
m_{11}&=&\frac{p_{11}^2N_{v,1}N_{h,1}b_1^2\beta_{vh}\beta_{hv}\nu_v + p_{21}^2 N_{v,1}N_{h,2}b_1^2\beta_{vh}\beta_{hv}\nu_v}{(p_{11}N_{h,1}+p_{21}N_{h,2})^2\gamma_h\mu_v(\nu_v+\mu_v)}\\
&=&\left( \frac{p_{11}^2N_{h,1}+p_{21}^2N_{h,2}}{(p_{11}N_{h,1}+p_{21}N_{h,2})^2}\right) \left( \frac{N_{v,1}b_1^2\beta_{vh}\beta_{hv}\nu_v}{\gamma_h\mu_v(\nu_v+\mu_v)}\right),
\end{eqnarray*}
\begin{eqnarray*}
m_{12}&=&\frac{p_{11}p_{12}N_{v,1}N_{h,1}b_1b_2\beta_{vh}\beta_{hv}\nu_v + p_{21}p_{22}N_{v,1}N_{h,2}b_1b_2\beta_{vh}\beta_{hv}\nu_v}{(p_{11}N_{h,1}+p_{21}N_{h,2})(p_{12}N_{h,1}+p_{22}N_{h,2})\gamma_h\mu_v(\nu_v+\mu_v)}\\
&=&\left( \frac{p_{11}p_{12}N_{h,1}+p_{21}p_{22}N_{h,2}}{(p_{11}N_{h,1}+p_{21}N_{h,2})(p_{12}N_{h,1}+p_{22}N_{h,2})}\right) \left( \frac{N_{v,1}b_1b_2\beta_{vh}\beta_{hv}\nu_v}{\gamma_h\mu_v(\nu_v+\mu_v)} \right),
\end{eqnarray*}
\begin{eqnarray*}
m_{21}&=&\frac{p_{11}p_{12}N_{v,2}N_{h,1}b_1b_2\beta_{vh}\beta_{hv}\nu_v + p_{21}p_{22}N_{v,2}N_{h,2}b_1b_2\beta_{vh}\beta_{hv}\nu_v}{(p_{11}N_{h,1}+p_{21}N_{h,2})(p_{12}N_{h,1}+p_{22}N_{h,2})\gamma_h\mu_v(\nu_v+\mu_v)}\\
&=&\left( \frac{p_{11}p_{12}N_{h,1}+p_{21}p_{22}N_{h,2}}{(p_{11}N_{h,1}+p_{21}N_{h,2})(p_{12}N_{h,1}+p_{22}N_{h,2})}\right) \left( \frac{N_{v,2}b_1b_2\beta_{vh}\beta_{hv}\nu_v}{\gamma_h\mu_v(\nu_v+\mu_v)} \right).
\end{eqnarray*}
\begin{eqnarray*}
m_{22}&=&\frac{p_{12}^2N_{v,2}N_{h,1}b_2^2\beta_{vh}\beta_{hv}\nu_v + p_{22}^2N_{v,2}N_{h,2}b_2^2\beta_{vh}\beta_{hv}\nu_v}{(p_{12}N_{h,1}+p_{22}N_{h,2})^2\gamma_h\mu_v(\nu_v+\mu_v)}\\
&=&\left( \frac{p_{12}^2N_{h,1} + p_{22}^2N_{h,2} }{(p_{12}N_{h,1}+p_{22}N_{h,2})^2}\right)  \left( \frac{N_{v,2}b_2^2\beta_{vh}\beta_{hv}\nu_v}{\gamma_h\mu_v(\nu_v+\mu_v)} \right).
\end{eqnarray*}
Particularly, $$\mathcal R_0^2=\frac{1}{2}\left(m_{11}+m_{22}+\sqrt{(m_{11}-m_{22})^2-4m_{12}m_{21}}\right)$$
We also have the relationship  $\frac{m_{12}}{m_{21}}=\frac{N_{v,1}}{N_{v,2}}$. If the two are isolated, that is, $p_{11} =p_{22} =1$, a case that allow us to estimate the power of Zv transmission in the absence of mobility, that is, when each community deals with Zv independently and mobility is not allowed, the basic reproduction number is $\mathcal R_0^2=\max \{\mathcal R_{0,1}^2, \mathcal R_{0,2}^2 \}$ where, for $i=1,2$,
  
$$\mathcal R_{0,i}^2=\frac{N_{v,i}b_i^2\beta_{vh}\beta_{hv}\nu_v}{N_{h,i}\gamma_h\mu_v(\nu_v+\mu_v)}$$

And so, we proceed to  make use of the disparity between patches assumption,  captured by the inequality, $\hat{\beta_1} \gg  \hat{\beta_2}$, where $\hat{\beta_i}$ is directly proportional to the local basic reproduction number or $\mathcal R_{0i}, i =1,2$, which defines the risk in patch $i, i=1,2$,  [high risk Patch 1 and  low risk Patch 2] under idealized, that is, no mobility conditions. Changes in  the entries of $\mathbb{P}$ lead to changes in the global basic reproduction number, $\mathcal R_0$, which naturally impact for example, the long-term dynamics of Zv in the two-patch system. It impacts the overall final epidemic size as well as the residents prevalence within each patch. A possible scenarios that allow us to explore the role that mobility, $\mathbb{P}$ and local risk, $\mathcal R_{0i}$ have on global risk ($\mathcal R_0$) and on the prevalence of infections among the residents of each patch are explored via simulations.

\section{Simulations}
Simulations are used to assess the impact of local mobility, on the dynamics of Zv, between two close communities. The first heavily affected by violence, poverty and lack of resources and a second with  access to public health vector control measures, health facilities, resources and the ability to minimize crime and violence. We assume the high risk community (Patch 1) is at a higher risk of acquiring Zv; this risk will be manifested through higher biting rates and a higher vector-host ratio leading well captured with a high reproductive number.

The simulations of this idealized world considers two scenarios for Patch 1, the first defined by a high  local basic reproductive number, $\mathcal{R}_{01}=3$ and the second by taking $\mathcal{R}_{01}=1.5$, both within the ranges of previous Zv outbreaks \cite{towers2016barranquilla,kucharski2016transmission}.  It is assumed that in the absence of mobility, Patch 2 would be unable to support a Zv outbreak and since its mobility-free reproductive number is assumed to be $\mathcal{R}_{02}=0.9$. It is further assumed that mobility from Patch 2 into Patch 1 is unappealing due to, for example,   high levels of violent  activity in Patch 1. And so, individuals from Patch 2 spend on average a a limited amount of their day in Patch 1; thus for our baseline simulations we take $p_{21}=0.10$.

\begin{figure}[H]
\centering
\includegraphics[scale=0.55]{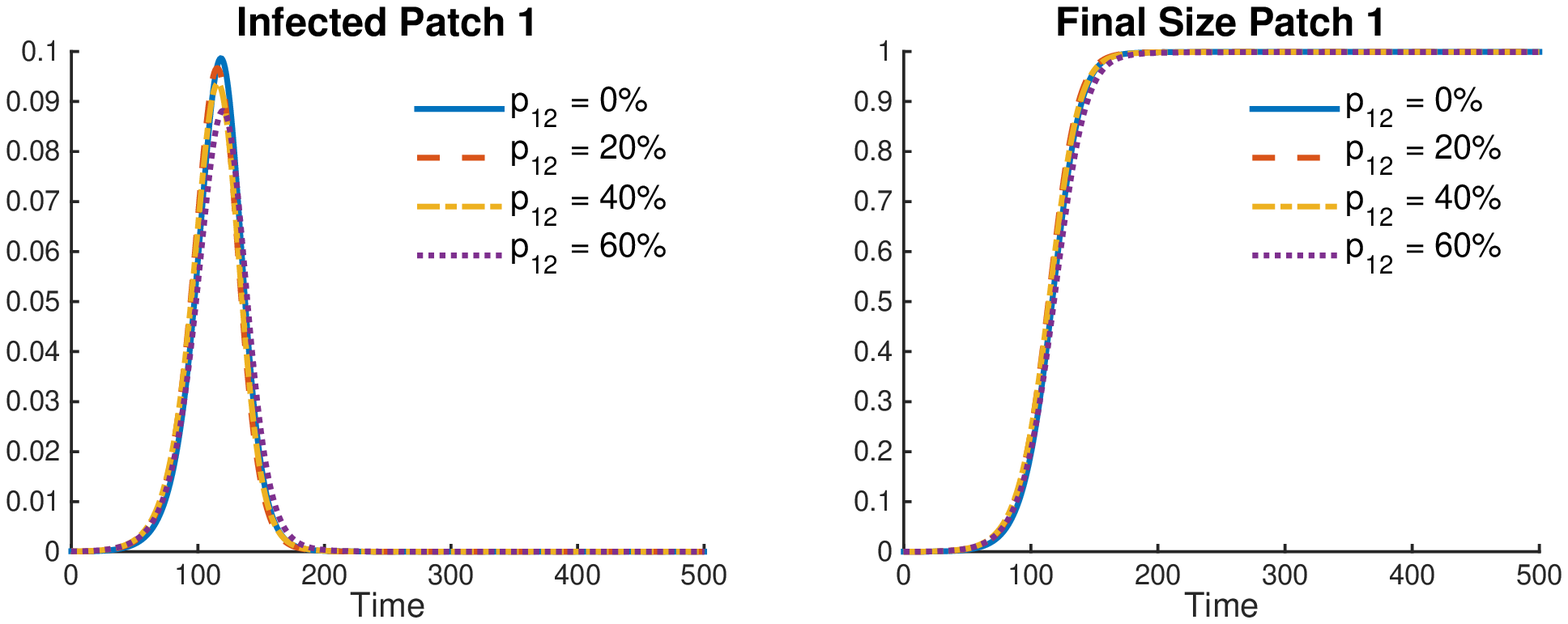}
\includegraphics[scale=0.55]{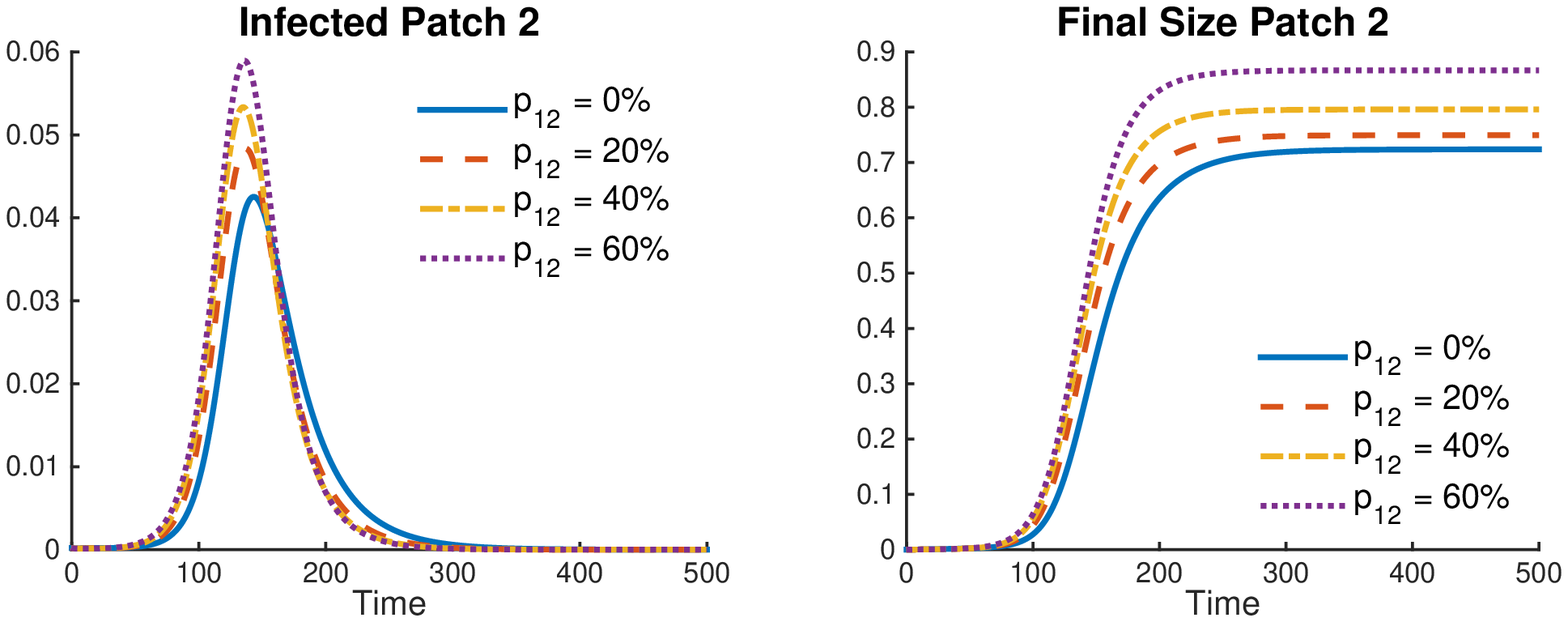}
\caption{Proportion of incidence and proportion of final infected individuals in Patch 1 and Patch 2 when $p_{21}=0.10$, $p_{12}=0, 0.20, 0.40$ and $0.60$. Mobility does not have a significant effect in the final size of Patch 1, while the final size of Patch 2 increases significantly. $\mathcal{R}_{01}=3$ and $\mathcal{R}_{02}=0.9$.}
\label{fig2}
\end{figure}

Figure \ref{fig2}, shows the proliferation of the outbreak as a result of mobility when we assume the reproduction number from the current Barranquilla outbreak ($\mathcal{R}_{01}=3$) \cite{towers2016barranquilla}. Both the incidence and final size of Patch 1 and Patch 2 increase for all the mobility values tested.

\begin{figure}[H]
\centering
\includegraphics[scale=0.55]{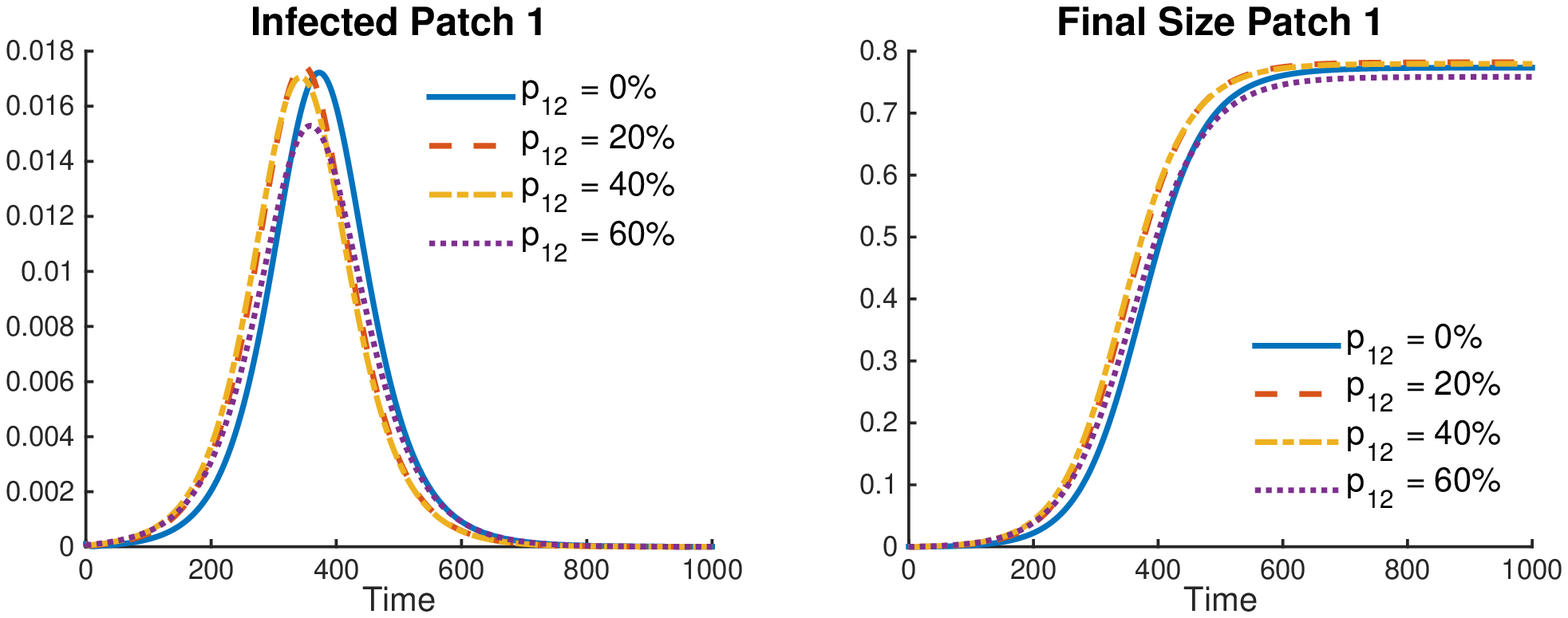}
\includegraphics[scale=0.55]{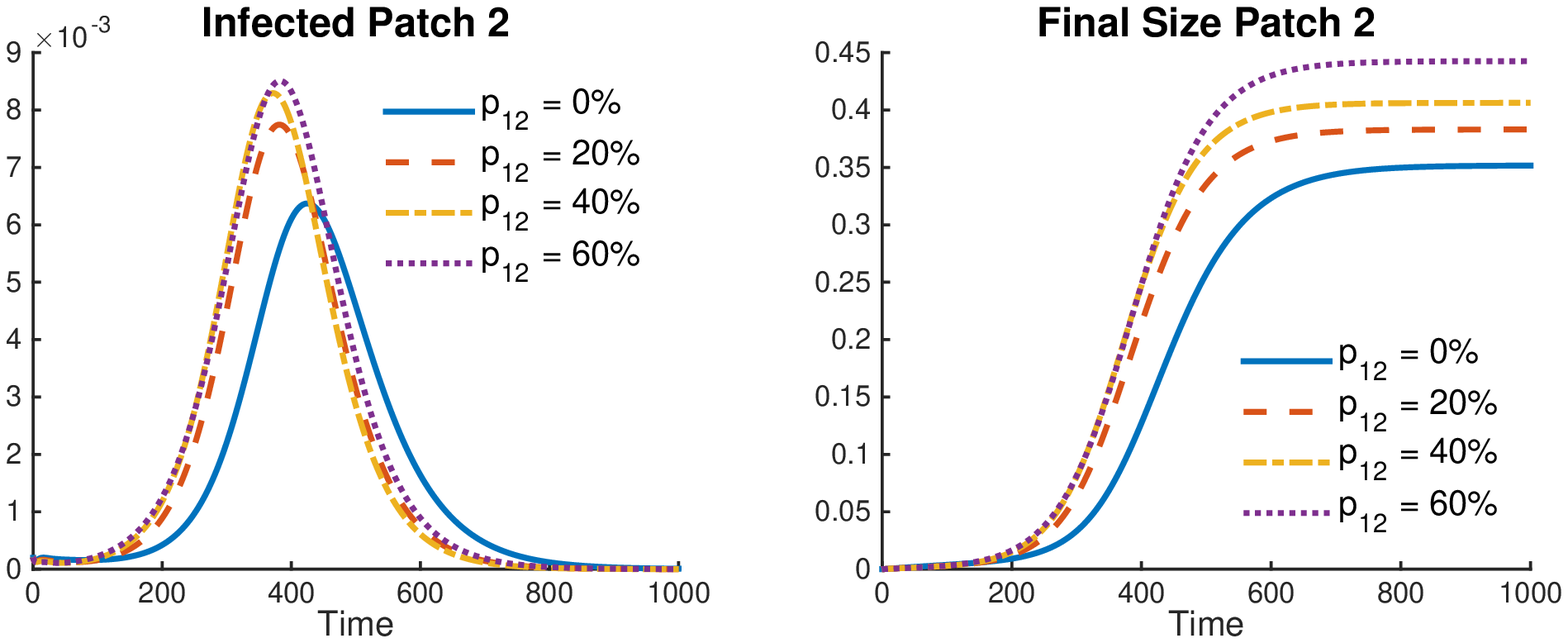}
\caption{Proportion of incidence and proportion of final infected individuals in Patch 1 and Patch 2, for traveling times $p_{21}=0.10$ and $p_{12}=0, 0.20, 0.40$ and $0.60$. There is no significant change in Patch 1, but a notable increment of the final size in Patch 2. $\mathcal{R}_{01}=1.5$ and $\mathcal{R}_{02}=0.9$.}
\label{fig1}
\end{figure}

Considering a a lower reproductive number, $\mathcal{R}_{01}=1.5$, Figure \ref{fig1} suggest that under high mobility values, the behavior of the final size in Patch 1 shifts, meaning that for particular mobility values (around 0.6 or greater) for which Patch 1 benefits from mobility. Nonetheless, this reduction on the final size of Patch 1 is not significant enough to have a positive effect on the outcome of the global final size.
%

\begin{figure}[H]
\centering
\includegraphics[scale=0.55]{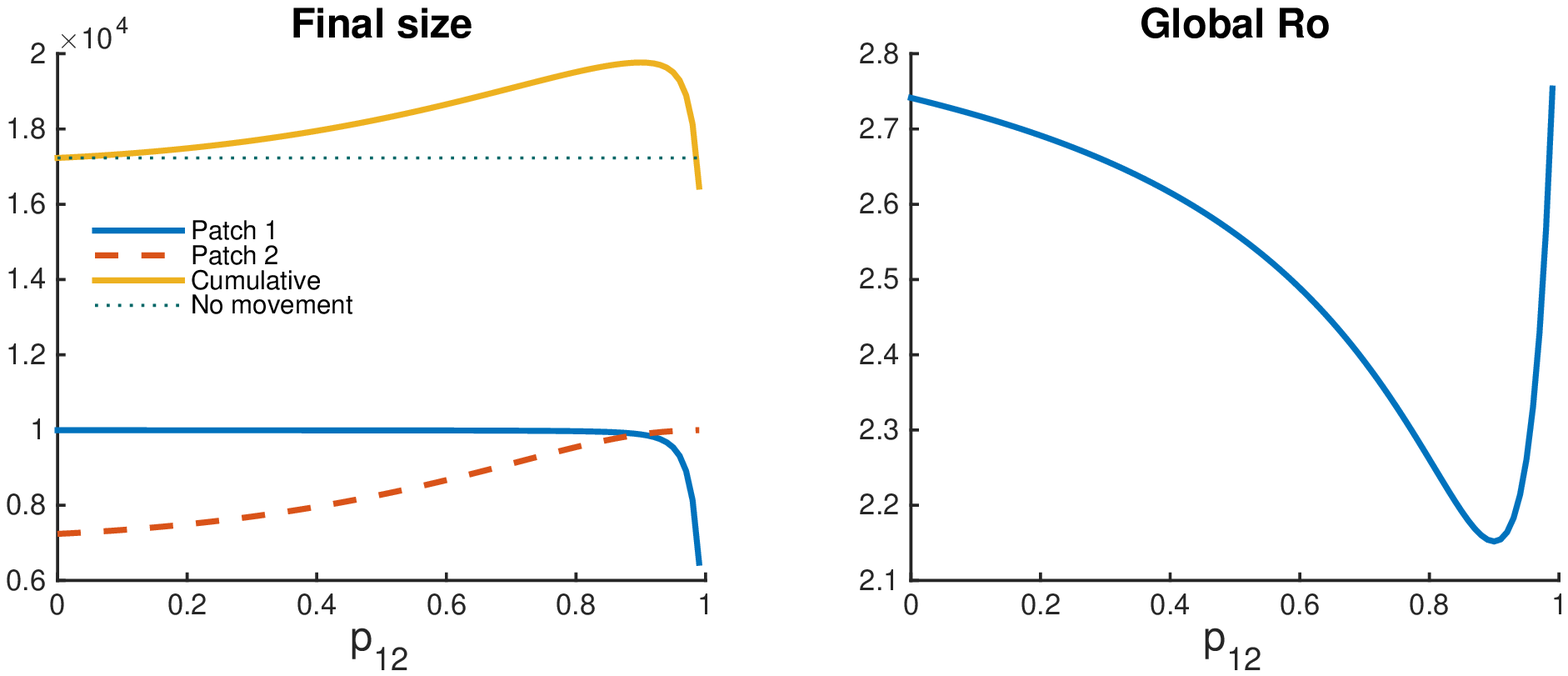}
\includegraphics[scale=0.55]{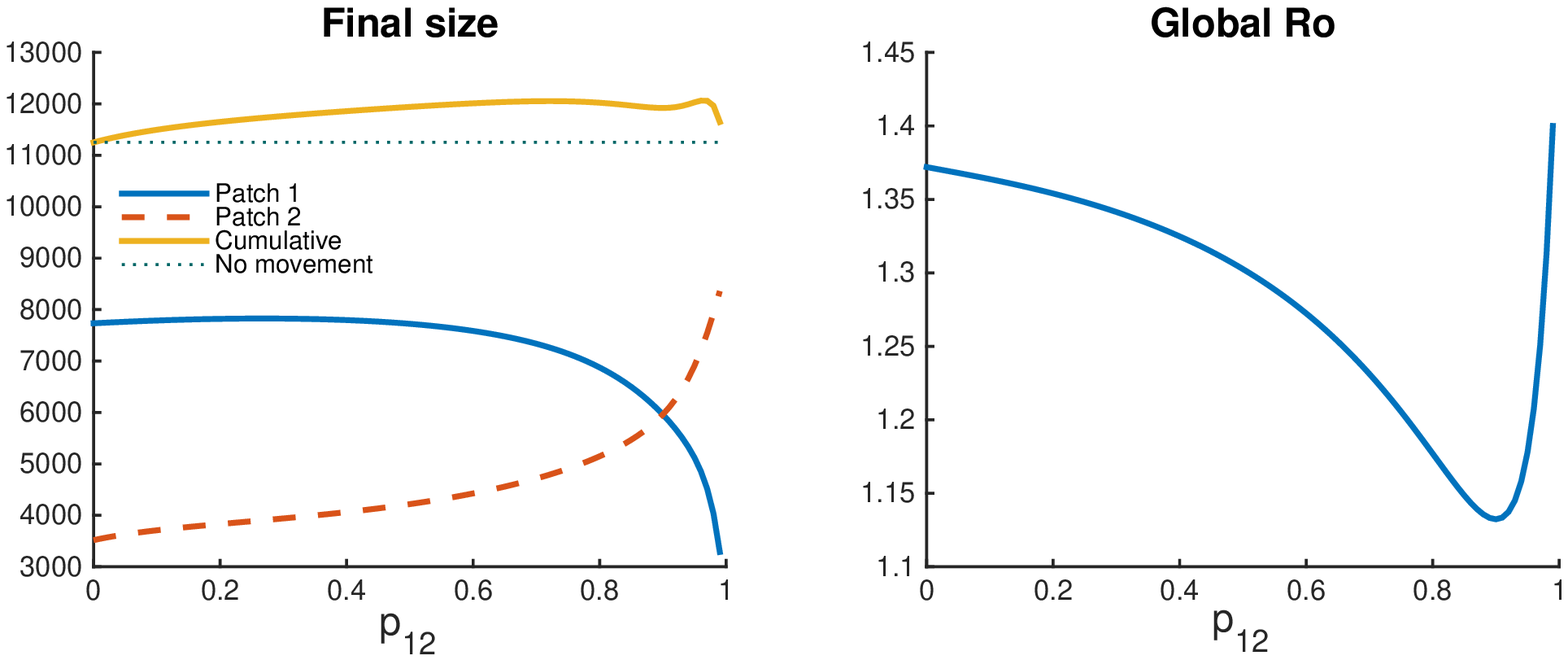}
\caption{Local and global final sizes through mobility values when $p_{21}=0.10$. Although mobility reduces the global $\mathcal{R}_0$, allowing mobility in the case of El Salvador ($\mathcal{R}_0=2$) leads to a detrimental effect in the global final size.}
\label{finsro}
\end{figure}

%
Analysis of the Final size as a function of residence times (mobility values) show, in Figure \ref{finsro}, the aforementioned behavior along with a change in behavior for the case where $\mathcal{R}_{01}=1.5$. Furthermore, the global reproductive number $\mathcal{R}_0$ is reduced for almost all mobility values compared to the zero mobility case. It is important to notice that the global reproductive number never drops below 1, meaning that for these two particular cases ($\mathcal{R}_{01}=1.5,3$), mobility is not enough to bring $\mathcal{R}_0<1$.
%

\begin{figure}[H]
\centering
\includegraphics[scale=0.45]{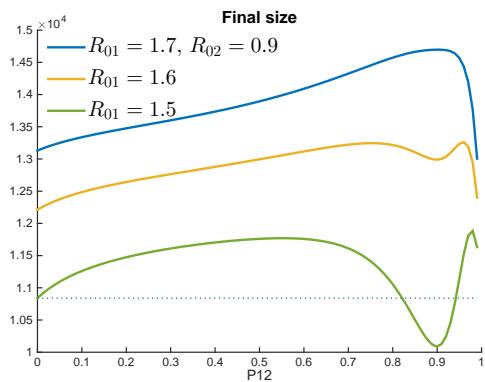}
\caption{Global final size through mobility values when $p_{21}=0.10$. $\mathcal{R}^2_0$ is fixed to 0.9, meanwhile $\mathcal{R}_{01}$ is varied. $\mathcal{R}_{01}=1.5$ shows an interval of residence times that reduces the final size.}
\label{ros1}
\end{figure}

Figure \ref{ros1}, suggests that  a small change in the local basic reproductive number of the high risk Patch, is able to drive a considerable change in the behavior of the cumulative final size. Meaning that the effective implementation of control measures along with specific mobility patters could have a beneficial impact on the outcome of the epidemic.

\begin{figure}[H]
\centering
\includegraphics[scale=0.4]{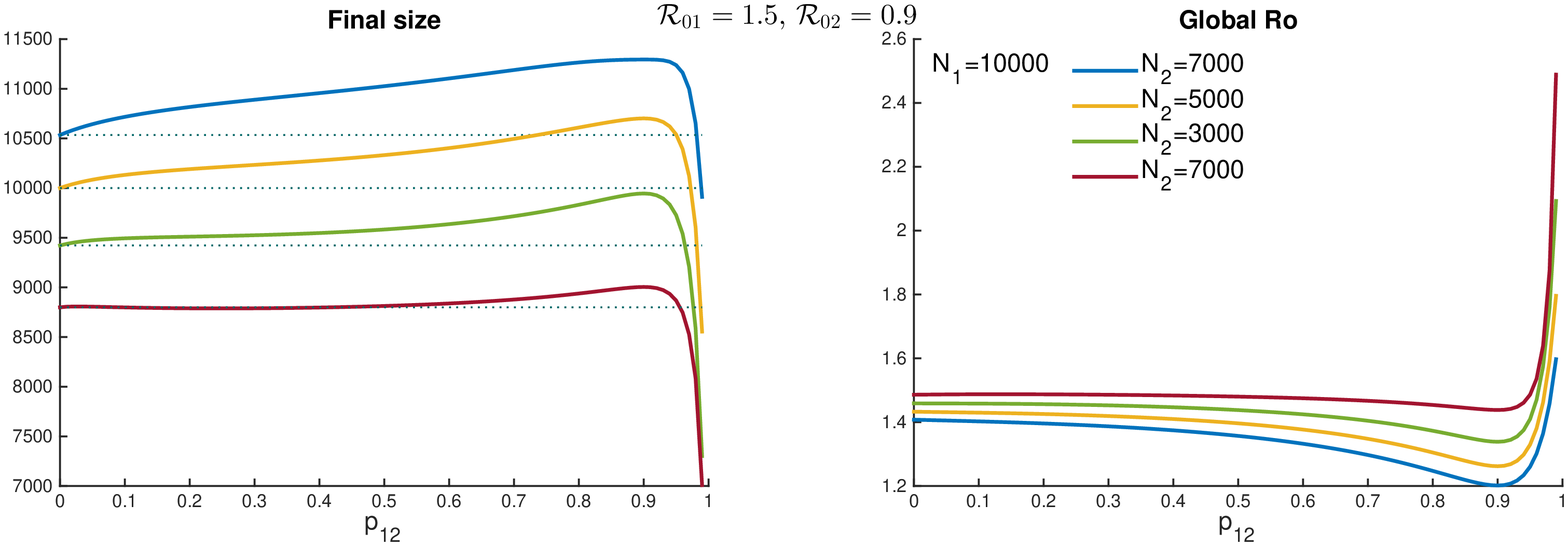}
\includegraphics[scale=0.4]{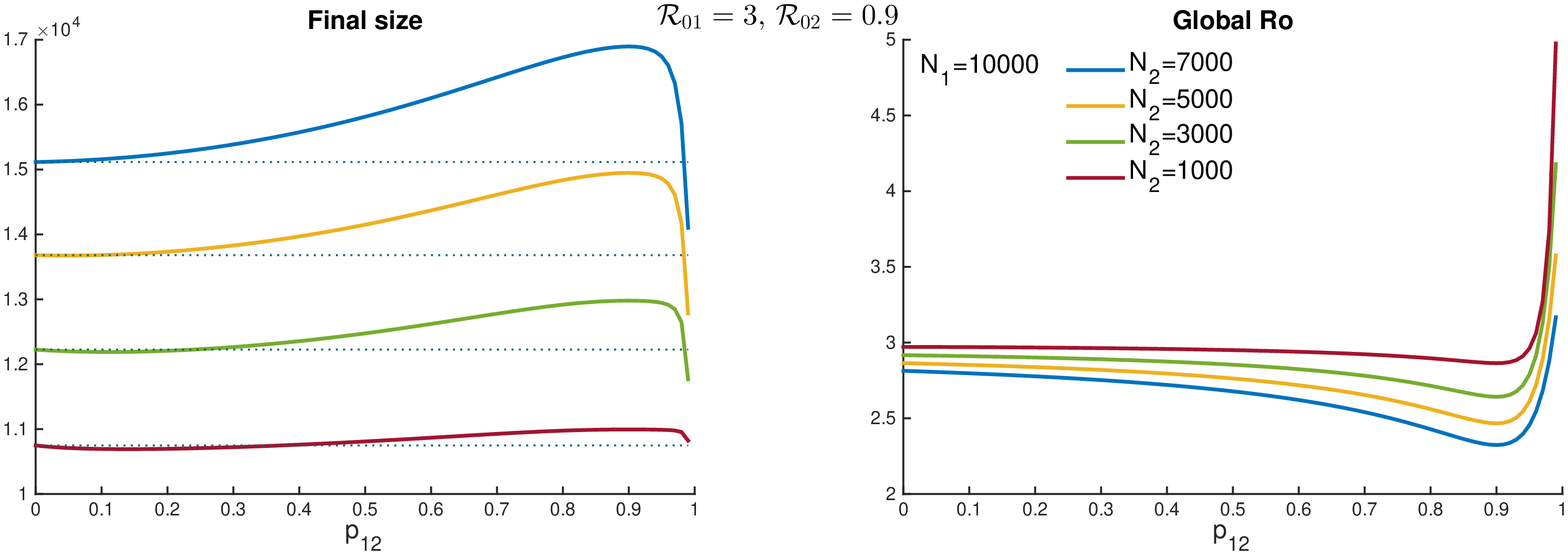}
\label{prop}
\caption{Cumulative final size and global $\mathcal{R}_0$ under mobility ($p_{21}=0.10$) and population proportions effects. When $N_1\gg N_2$, it is possible to find a set of residence times (around $0.20$) that reduces the cumulative final size in comparison with the zero movement case}
\label{ros}
\end{figure}

Finally, we wanted to study the effect produced by the population size from Patch 2 in the cumulative final size and the global $\mathcal{R}_0$. Figure \ref{ros} summarize the mobility effects for both scenarios $\mathcal{R}_{01}=3$ and $\mathcal{R}_{01}=1.5$, while $\mathcal{R}_{02}=0.9$ and $N_1=10000$. In both scenarios, the effect of mobility is adverse when $N_1\approx N_2$,  however when $N_1\gg N_2$, it is possible to find a set of residence times (around $0.20$) that reduces the cumulative final size in comparison with the zero movement case.
On the other hand, the global $\mathcal{R}_0$ is being reduced considerably when $N_1\approx N_2$, while for the case $N_1\gg N_2$ the global $\mathcal{R}_0$ remains almost equal.

For the two epidemiological scenarios $\mathcal{R}_{01}=3$ and $\mathcal{R}_{01}=1.5$,  tables \ref{tab:prop1} and \ref{tab:prop2} show a summary of the average proportion of infected population when low ($p_{12}=0-0.32$), intermediate ($p_{12}=0.33- 0.65$) and high mobility ($p_{12}=0.66- 0.99$) is allowed for $p_{21}=0.10$. A fixed population size for Patch 1 $N_1=10000$ and different population sizes for Patch 2 are considered $N_2=1000,3000,5000,7000$ and the case $N_1=N_2=10000$.
\begin{table}[H]
  \begin{center}
    \caption{Final size (Patch 1, Patch 2) $N_1=10000$, $\mathcal{R}_{01}=3$, $\mathcal{R}_{02}=0.9$ and $p_{21}=0.10$.}
    \label{tab:prop1}
    \begin{tabular}{ccccc}
      \cline{1-5}
      $N_2$ & Low Mobility & Intermediate Mobility & High Mobility & Min $\mathcal{R}_0$ \\
      \cline{1-5}
1000 &(0.9996, 0.7110) &(0.9996, 0.8124) &(0.9987, 0.9757) & 2.8643 \\
3000 &(0.9996, 0.7398) &(0.9995, 0.8263) &(0.9935, 0.9722) & 2.6421 \\
5000 &(0.9995, 0.7468) &(0.9994, 0.8308) &(0.9870, 0.9688) & 2.4666 \\
7000 &(0.9994, 0.7478) &(0.9992, 0.8310) &(0.9807, 0.9648) & 2.3237 \\
10000  &(0.9992, 0.7451) &(0.9989, 0.8276) &(0.9720, 0.9579) & 2.1519 \\
      \cline{1-5}
    \end{tabular}
  \end{center}
\end{table}

\begin{figure}[H]
\centering
\includegraphics[scale=0.4]{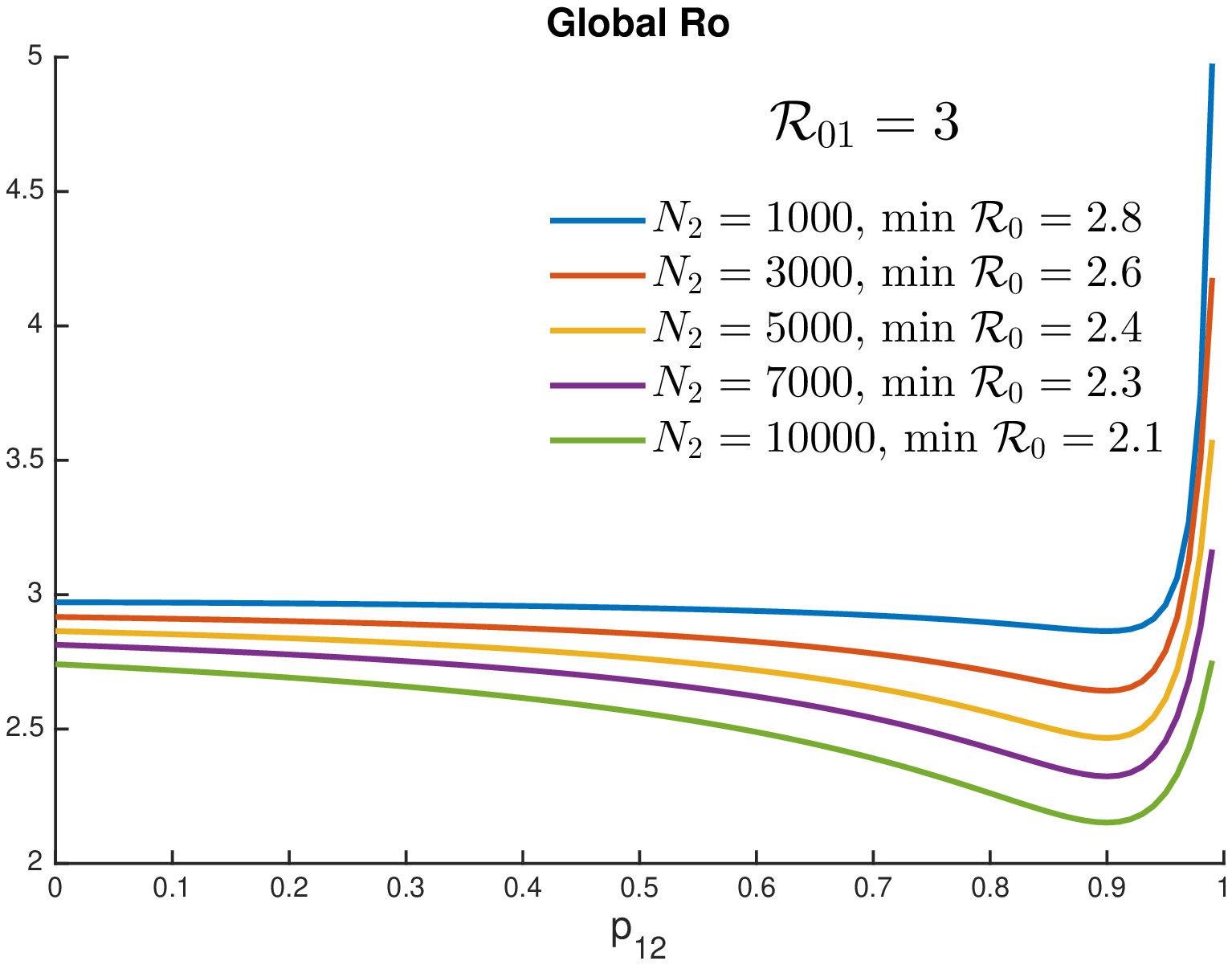}\hspace*{0.3cm}\includegraphics[scale=0.4]{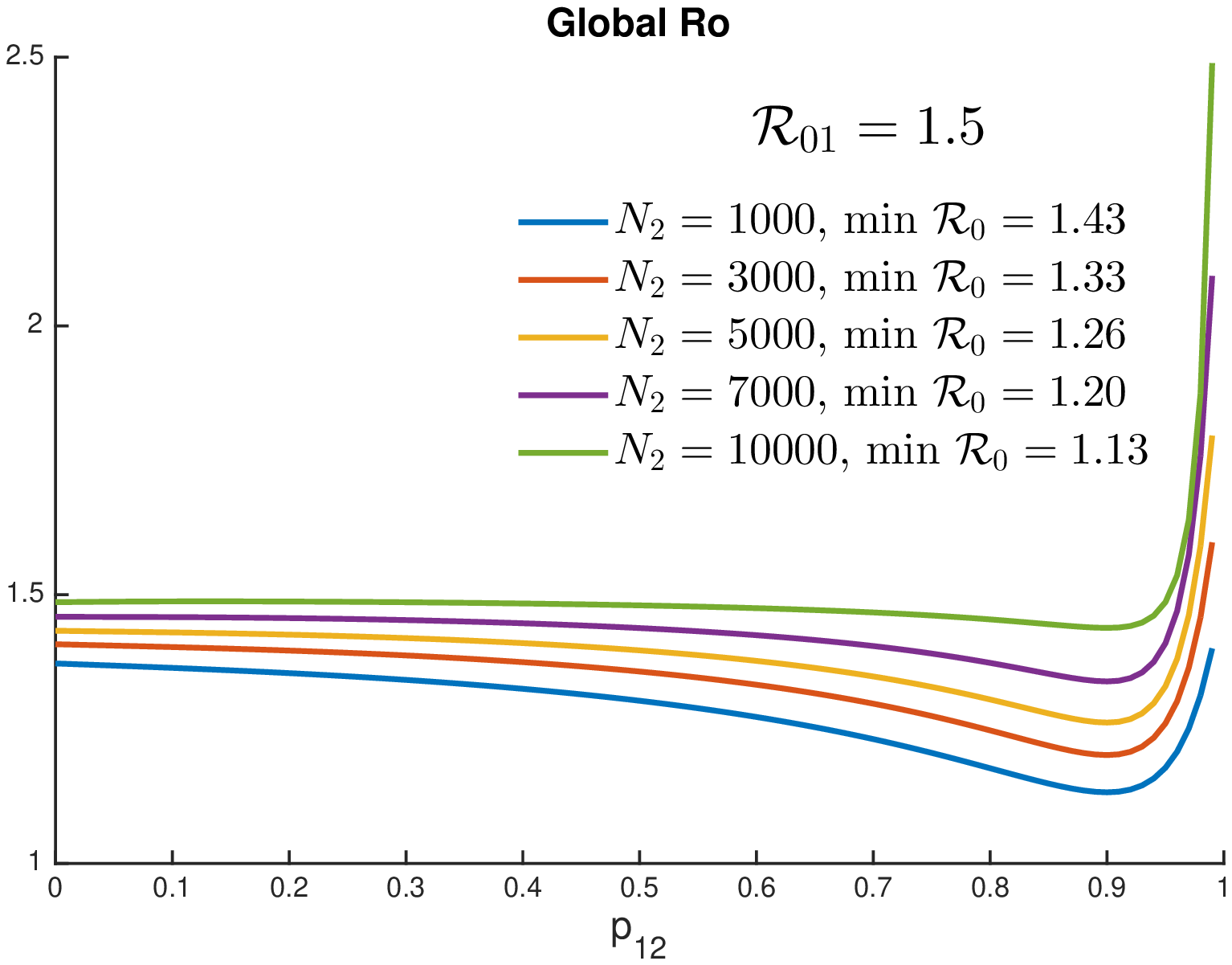}
\caption{Global $\mathcal{R}_0$ dynamics through mobility  when $p_{21}=0.10$. Patch 2 populations are varied from $N_2=1000,3000,5000,7000$ up to $10000$. The global $\mathcal{R}_0$ hits its minimum always at $90\%$ of mobility and as $N_1\approx N_2$ this minimum value is decreasing.}
\label{rodata}
\end{figure}

\begin{table}[H]
  \begin{center}
    \caption{Final size (Patch 1, Patch 2) $N_1=10000$, $\mathcal{R}_{01}=1.5$, $\mathcal{R}_{02}=0.9$ and $p_{21}=0.10$.}
    \label{tab:prop2}
    \begin{tabular}{ccccc}
      \cline{1-5}
      $N_2$ & Low Mobility & Intermediate Mobility & High Mobility & Min $\mathcal{R}_0$ \\
      \cline{1-5}
1000 &(0.8442, 0.3526) &(0.8442, 0.3736) &(0.8144, 0.6902) & 1.4382 \\
3000 &(0.8332, 0.3886) &(0.8336, 0.4169) &(0.7673, 0.6702) & 1.3386 \\
5000 &(0.8194, 0.3924) &(0.8183, 0.4299) &(0.7272, 0.6427) & 1.262 \\
7000 &(0.8043, 0.3883) &(0.8005, 0.4310) & (0.6899, 0.6117) & 1.2016\\
10000  &(0.7799, 0.3771) &(0.7704, 0.4223) &(0.6367, 0.5630) & 1.1323 \\
      \cline{1-5}
    \end{tabular}
  \end{center}
\end{table}
Figure \ref{rodata} shows the global $\mathcal{R}_0$ over all mobility values for different population sizes of Patch 2, for the two epidemic scenarios. The minimum $\mathcal{R}_0$ value is  reached for all cases when mobility is at $90\%$ and this value is being reduced when $N_1\approx N_2$. The global $\mathcal{R}_0$ is dominated by the local $\mathcal{R}_{01}$, this is, the local $\mathcal{R}_0$ of the high risk Patch. 

\section{Conclusions}
This manuscript looks at the role of mobility in an idealized setting involving two adjacent highly distinct communities. The first community has the resources and means to control a Zv outbreak ($\mathcal{R}_{02}<1$) while the second faces dramatic limitations, $\mathcal{R}_{01}$, very large, or strong limitations $\mathcal{R}_{01}$, large.  We also explored the role of density by assuming that $N_1= k N_2$ with k =1,..,10. Each simulation looks at the role of the mobility matrix $\mathbb{P}$, the global $\mathcal{R}_{0}$, and local $\mathcal{R}_{0i}$, i=1,2. On the local and overall prevalence of Zv. We verify the expected results, density matters, the global $\mathcal{R}_{0}$ has a minimum value, which depends on the entries of the mobility matrix 
$\mathbb{P}$, and $\mathcal{R}_{01}$, very large is a lot worse than $\mathcal{R}_{01}$ large. We also see that unless movement is dramatically reduced that there is no hope that Zv can be contained. Certainly, stopping mobility would lead to the elimination of Zv in Patch 2. By stopping mobility would bring the economy to a halt, in this idealized two-world community.  Naturally, the set up is not representative of any real situation. The model could be easily modified to include the type of heterogeneity found in `real' communities. The dynamics of Zv in places where violence is high due to gangs or other type of criminal activities, would be no doubt, make it nearly impossible to eliminate Zv.

\section{Acknowledgements}
This project has been partially supported by grants from the National Science Foundation (DMS-1263374 and DUE-1101782), the National Security Agency (H98230-14-1-0157), the Office of the President of ASU, and the Office of the Provost of ASU. The views expressed are sole responsibility of the authors and not the funding agencies.



\begin{thebibliography}{10}

\bibitem{bichara2015vector}
{\sc D.~Bichara and C.~Castillo-Chavez}, {\em Vector-borne diseases models with
  residence times-a lagrangian perspective}, arXiv preprint arXiv:1509.08894,
  (2015).

\bibitem{bichara2016dynamics}
{\sc D.~Bichara, S.~A. Holecheck, J.~Velazquez-Castro, A.~L. Murillo, and
  C.~Castillo-Chavez}, {\em On the dynamics of dengue virus type 2 with
  residence times and vertical transmission}, arXiv preprint arXiv:1601.06234,
  (2016).

\bibitem{bichara2015sis}
{\sc D.~Bichara, Y.~Kang, C.~Castillo-Chavez, R.~Horan, and C.~Perrings}, {\em
  Sis and sir epidemic models under virtual dispersal}, Bulletin of
  mathematical biology, 77 (2015), pp.~2004--2034.

\bibitem{cao2014emerging}
{\sc V.-M. Cao-Lormeau and D.~Musso}, {\em Emerging arboviruses in the
  pacific}, The Lancet, 384 (2014), pp.~1571--1572.

\bibitem{is2014zika}
{\sc V.~M. Cao-Lormeau, C.~Roche, A.~Teissier, E.~Robin, A.-L. Bery, H.-P.
  Mallet, A.~A. Sall, and D.~Musso}, {\em Zika virus, french polynesia, south
  pacific, 2013},  (2014).

\bibitem{CDC2016e}
{\sc CDC(\lowercase{a})}, {\em Zika virus}.
\newblock Online, February 18, 2016 2016.

\bibitem{CDC2016}
\leavevmode\vrule height 2pt depth -1.6pt width 23pt, {\em Zika virus}.
\newblock Online, February 01, 2016 2016.

\bibitem{CDC2016b}
{\sc CDC(\lowercase{b})}, {\em \uppercase{CDC} adds 2 destinations to interim
  travel guidance related to zika virus}.
\newblock Online, February 2016.

\bibitem{CDC2016c}
{\sc CDC(\lowercase{c})}, {\em Zika virus in \uppercase{S}outh
  \uppercase{A}merica}.
\newblock Online, February 2016.

\bibitem{CDC2016d}
{\sc CDC(\lowercase{d})}, {\em Zika virus in \uppercase{C}entral
  \uppercase{A}merica}.
\newblock Online, February 2016.

\bibitem{dick1952zika}
{\sc G.~Dick, S.~Kitchen, and A.~Haddow}, {\em Zika virus (i). isolations and
  serological specificity}, Transactions of the Royal Society of Tropical
  Medicine and Hygiene, 46 (1952), pp.~509--520.

\bibitem{duffy2009zika}
{\sc M.~R. Duffy, T.-H. Chen, W.~T. Hancock, A.~M. Powers, J.~L. Kool, R.~S.
  Lanciotti, M.~Pretrick, M.~Marfel, S.~Holzbauer, C.~Dubray, et~al.}, {\em
  Zika virus outbreak on yap island, federated states of micronesia}, New
  England Journal of Medicine, 360 (2009), pp.~2536--2543.

\bibitem{dupont2015co}
{\sc M.~Dupont-Rouzeyrol, O.~O'Connor, E.~Calvez, M.~Daures, M.~John, J.-P.
  Grangeon, and A.-C. Gourinat}, {\em Co-infection with zika and dengue viruses
  in 2 patients, new caledonia, 2014}, Emerging infectious diseases, 21 (2015),
  p.~381.

\bibitem{fauci2016zika}
{\sc A.~S. Fauci and D.~M. Morens}, {\em Zika virus in the americas---yet
  another arbovirus threat}, New England Journal of Medicine,  (2016).

\bibitem{faye2014molecular}
{\sc O.~Faye, C.~C. Freire, A.~Iamarino, O.~Faye, J.~V.~C. de~Oliveira,
  M.~Diallo, P.~M. Zanotto, et~al.}, {\em Molecular evolution of zika virus
  during its emergence in the 20 th century}, PLoS Negl Trop Dis, 8 (2014),
  p.~e2636.

\bibitem{haddow2012genetic}
{\sc A.~D. Haddow, A.~J. Schuh, C.~Y. Yasuda, M.~R. Kasper, V.~Heang, R.~Huy,
  H.~Guzman, R.~B. Tesh, and S.~C. Weaver}, {\em Genetic characterization of
  zika virus strains: geographic expansion of the asian lineage}, PLoS Negl
  Trop Dis, 6 (2012), p.~e1477.

\bibitem{hayes2009zika}
{\sc E.~B. Hayes et~al.}, {\em Zika virus outside africa}, Emerg Infect Dis, 15
  (2009), pp.~1347--1350.

\bibitem{BID2015}
{\sc L.~Jaitman}, {\em Los costos del crimen y la violencia en el bienestar en
  America Latina y el Caribe, Laura Jaitman Ed.}, Banco Interamericano del
  Desarrollo, 2015.

\bibitem{kucharski2016transmission}
{\sc A.~J. Kucharski, S.~Funk, R.~M. Eggo, H.-P. Mallet, J.~Edmunds, and E.~J.
  Nilles}, {\em Transmission dynamics of zika virus in island populations: a
  modelling analysis of the 2013-14 french polynesia outbreak}, bioRxiv,
  (2016), p.~038588.

\bibitem{macnamara1954zika}
{\sc F.~Macnamara}, {\em Zika virus: a report on three cases of human infection
  during an epidemic of jaundice in nigeria}, Transactions of the Royal Society
  of Tropical Medicine and Hygiene, 48 (1954), pp.~139--145.

\bibitem{musso2015zika}
{\sc D.~Musso, V.~M. Cao-Lormeau, and D.~J. Gubler}, {\em Zika virus: following
  the path of dengue and chikungunya?}, The Lancet, 386 (2015), pp.~243--244.

\bibitem{musso2014rapid}
{\sc D.~Musso, E.~Nilles, and V.-M. Cao-Lormeau}, {\em Rapid spread of emerging
  zika virus in the pacific area}, Clinical Microbiology and Infection, 20
  (2014), pp.~O595--O596.

\bibitem{petersen2016interim}
{\sc E.~E. Petersen}, {\em Interim guidelines for pregnant women during a zika
  virus outbreak---united states, 2016}, MMWR. Morbidity and mortality weekly
  report, 65 (2016).

\bibitem{salvador2015entry}
{\sc F.~S. Salvador and D.~M. Fujita}, {\em Entry routes for zika virus in
  brazil after 2014 world cup: New possibilities}, Travel medicine and
  infectious disease,  (2015).

\bibitem{towers2016barranquilla}
{\sc S.~Towers, F.~Brauer, C.~Castillo-Chavez, A.~K. Falconar, A.~Mubayi, and
  C.~M. Romero-Vivas}, {\em Estimation of the reproduction number of the 2015
  zika virus outbreak in barranquilla, colombia}, Submitted.

\bibitem{WHO2016}
{\sc W.~H.~O. (WHO)}, {\em Zika virus, fact sheet}.
\newblock Online, January 2016.

\bibitem{zanluca2015first}
{\sc C.~Zanluca, V.~C. A.~d. Melo, A.~L.~P. Mosimann, G.~I. V.~d. Santos, C.~N.
  D.~d. Santos, and K.~Luz}, {\em First report of autochthonous transmission of
  zika virus in brazil}, Mem{\'o}rias do Instituto Oswaldo Cruz, 110 (2015),
  pp.~569--572.

\end{thebibliography}
\end{document}